\documentclass[journal]{IEEEtran}

\usepackage{amsmath,graphicx}
\usepackage{amssymb,bm}

\begin{document}

%
%

\title{Quantization Noise Shaping for Information Maximizing ADCs}

\author{Arthur~J.~Redfern~and~Kun~Shi%
\thanks{A. Redfern and K. Shi are with the Texas Instruments Systems and Applications R\&D Center, Dallas, TX 75243 USA (e-mail: \{redfern, k-shi\}@ti.com).}}



\maketitle

%
%

\begin{abstract}
ADCs sit at the interface of the analog and digital worlds and fundamentally determine what information is available in the digital domain for processing.  This paper shows that a configurable ADC can be designed for signals with non constant information as a function of frequency such that within a fixed power budget the ADC maximizes the information in the converted signal by frequency shaping the quantization noise.  Quantization noise shaping can be realized via loop filter design for a single channel delta sigma ADC and extended to common time and frequency interleaved multi channel structures.  Results are presented for example wireline and wireless style channels.
\end{abstract}

%
%

\begin{IEEEkeywords}
ADC, quantization noise, shaping.
\end{IEEEkeywords}

%
%

%
\IEEEpeerreviewmaketitle

%
%

\section{Introduction}
\label{SectionIntroduction}

\IEEEPARstart{I}{t's} common for electronic devices to operate with constrained power budgets.  Within these devices ADCs sit at the interface of the analog and digital worlds and fundamentally determine what information is available in the digital domain for processing.  Sampling at a high frequency with a high number of bits allows a sizable contiguous block of frequencies to be reproduced with high fidelity, but also has the drawback of requiring a large amount of ADC power $P_{\text{ADC}} \propto \Delta f2^b$ where $\Delta f$ is the bandwidth and $b$ is the number of bits \cite{REF5}, \cite{REF6}.

\begin{figure}[htb]
\centering
\includegraphics[width=8.6cm]{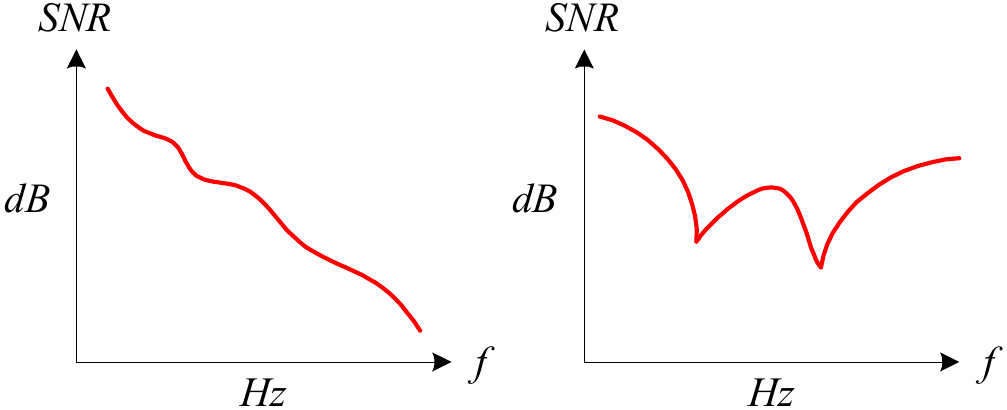}
\caption{Received signal SNR as a function of frequency for an example wireline (left) and wireless (right) channel.}
\label{FigureSignalSnr}
\end{figure}

Considering the analog signal in more detail, there are cases where the signal resides within a contiguous band of frequencies but within those frequencies the information content of the signal is non constant.  As an example, consider a multicarrier wireline or wireless communication system with bit loading where large constellation sizes are used in high SNR regions and small constellation sizes are used in low SNR regions (Fig.\ \ref{FigureSignalSnr}).

Traditionally, an ADC for this type of system would be designed with a number of bits capable of supporting the largest constellation size across the entire band.  However, this is power inefficient in the low SNR regions as many more bits are resolvable than the information content of the signal.  Likewise, the impact of the quantization noise is a nonuniform degradation of the received signal SNR, as an equivalent amount of quantization noise added to a high SNR region results in a larger degradation of SNR than if it was added to a low SNR region.

To address this, various ADCs have been proposed which allow for shaping the ADC quantization noise and thus the bits vs.\ frequency profile of the ADC.  For example, multi channel ADCs in the literature have shown how shaping can be done via changing the constant bits vs.\ frequency profile for delta sigma ADCs on a per channel basis \cite{REF2}, projecting the received signal on a basis optimized for the signal before conversion \cite{REF3}, \cite{REF8} and by allocating more ADCs to bands where the signal has a higher variance \cite{REF9}.  Efficient shaping for these cases relies on having a sufficiently large number of ADCs such that the SNR of the received signal does not change significantly within the band converted by the 1 of the individual constant bit vs.\ frequency ADCs which comprise the multi channel ADC structure.

Compressive sensing ADCs provide an implicit example of bits vs.\ frequency shaping for cases when the input signal is sparse in frequency and the sampling rate is proportional to the occupied signal bandwidth rather than the total system bandwidth \cite{REF4}.  Ignoring noise folding issues which arise when unwanted signals are also present, this can be viewed as an on/off shaping of the quantization noise where 0 bits are assigned to frequencies where there is no signal and a constant number of bits are assigned to frequencies where there is a signal.  As in the multi channel case, this can be viewed as block constant frequency shaping.

Given the existence of signals with information content which varies with frequency and ADCs which can be designed explicitly or implicitly for quantization noise shaping, the question arises as to what is the optimal quantization noise shape for an ADC with a fixed power budget to maximize the information content in the converted signal.  The key theoretical result in Section \ref{SectionInformation} is the derivation of an equation which answers this question and is independent of a specific ADC architecture.  Section \ref{SectionSingle} then connects this theory to a common ADC design by showing that quantization noise shaping can be achieved through the design of the loop filter in a delta sigma ADC.  Optimal quantization noise shaping is extended to time and frequency interleaved multi channel ADC structures in Section \ref{SectionMulti} and conclusions are provided in Section \ref{SectionConclusions}.

%
%

\section{Information maximization}
\label{SectionInformation}

The purpose of this section is to determine the ADC quantization noise shape that maximizes the information in the signal after the ADC.  Before the ADC, when the signal and noise are uncorrelated and the noise is additive colored Gaussian, the maximum information in a signal occupying frequencies $f_{A}$ to $f_{B}$ is
\begin{equation}
C_{b} = \int_{f_{A}}^{f_{B}}\log_{2}\left[1+\frac{S_{x}(f)}{S_{v}(f)}\right]df
\label{EqInfoBefore}
\end{equation}
where $f$ is frequency, $S_{x}(f)$ is the signal PSD and $S_{v}(f)$ is the noise PSD \cite{REF0}.

Modeling the effect of the ADC as adding shaped quantization noise with PSD $S_{q}(f)$ to the signal, the maximum information in the signal after the ADC is
\begin{equation}
C_{a} \approx \int_{f_{A}}^{f_{B}}\log_{2}\left[1+\frac{S_{x}(f)}{S_{v}(f) + S_{q}(f)}\right]df
\label{EqInfoAfter}
\end{equation}
where the approximation is due to the quantization noise having a uniform PDF and signal correlation.

The loss of information due to the ADC is found by subtracting (\ref{EqInfoAfter}) from (\ref{EqInfoBefore})
\begin{equation}
C_{\Delta}\!=\!C_{b}\!-\!C_{a}\approx\int_{f_{A}}^{f_{B}}\log_{2}\left[1+\frac{S_{q}(f)}{S_{v}(f)}\right]df
\label{EqInfoLoss}
\end{equation}
and assuming that the noise PSDs $S_{q}(f)$ and $S_{v}(f)$ are small relative to the signal PSD $S_{x}(f)$.

While small, the quantization noise is not arbitrarily small or 0 because the ADC is limited in power.  The quantization noise PSD and number of bits are related by
\begin{equation}
S_{q}(f) = 2^{-2b(f)}/12
\label{EqQnoise}
\end{equation}
and the ADC power and number of bits are related by
\begin{equation}
P_{\text{ADC}}=\frac{1}{c}\int_{f_{A}}^{f_{B}}2^{b(f)}df,
\label{EqPower}
\end{equation}
where $c$ is a proportionality constant that for convenience we can absorb in the definition of $P \equiv cP_{\text{ADC}}$.  Using (\ref{EqQnoise}) and (\ref{EqPower}) the quantization noise PSD and the ADC power are related as
\begin{equation}
\int_{f_{A}}^{f_{B}}S_{q}^{-\frac{1}{2}}(f)=\sqrt{12}P.
\label{EqPowerConstraint}
\end{equation}
The smaller the quantization noise, the larger the power of the ADC.

To determine the optimal quantization noise PSD shape which minimizes the information loss of the signal after the ADC (\ref{EqInfoLoss}) given the power constraint (\ref{EqPowerConstraint}), integrals are converted into Riemann sums by dividing the band from $f_{A}$ to $f_{B}$ into $K$ subchannels of bandwidth $(f_{B} - f_{A})/K$ indexed by $k = 1, \ldots,  K$ and forming the Lagrangian
\begin{align}
J[\lambda,S_{q}(k)]=&\frac{f_{B}-f_{A}}{K}\sum_{k=1}^{K}\log_{2}\left[1+\frac{S_{q}(k)}{S_{v}(k)}\right]\nonumber\\
&+\lambda\left(\frac{1}{K}\sum_{k=1}^{K}S_{q}^{-\frac{1}{2}}(k)-\frac{\sqrt{12}P}{f_{B}-f_{A}}\right),\label{EqLagrangian}
\end{align}
where $\lambda$ is a Lagrange multiplier.  As both the information loss (\ref{EqInfoLoss}) and the power constraint formed from (\ref{EqPowerConstraint}) are convex, their sum (\ref{EqLagrangian}) is also convex \cite{REF1}.

Taking first order partial derivatives with respect to $S_{q}(k)$ and $\lambda$, setting the results to 0 and using the assumption that $S_{q}(f)$ is small relative $S_{v}(f)$ creates the system of equations
\begin{align}
\frac{\partial{J}}{\partial{S_{q}(k)}}\!=\!0&\Rightarrow
S_{q}^{-\frac{1}{2}}\!(k)\!\approx\!\frac{2(f_{B}\!-\!f_{A})\log_{2}(e)}{\lambda}\frac{S_{q}(k)}{S_{v}(k)},\label{EqDerivativeSq}\\
\frac{\partial{J}}{\partial{\lambda}}\!=\!0&\Rightarrow\frac{1}{K}\sum_{k=1}^{K}S_{q}^{-\frac{1}{2}}(k)=\frac{\sqrt{12}P}{f_{B}-f_{A}}.\label{EqDerivativeLambda}
\end{align}
Substituting (\ref{EqDerivativeSq}) into (\ref{EqDerivativeLambda}), solving for $\lambda$, then substituting the result into (\ref{EqDerivativeSq}) and solving for $S_{q}(k)$ results in
\begin{align}
S_{q}(k)=S_{v}^{\frac{2}{3}}(k)\left[\frac{\frac{f_{B}-f_{A}}{K}\sum_{k=1}^{K}\frac{S_{q}(k)}{S_{v}(k)}}{\sqrt{12}P}\right]^{\frac{2}{3}}.
\label{EqQuantSum}
\end{align}

While (\ref{EqQuantSum}) relates $S_{q}(k)$ to $S_{v}(k)$, it's somewhat cumbersome to use as $S_{q}(k)$ occurs on both sides of the equation.  To get rid of the summation term with $S_{q}(k)$ on the right hand side form an equivalent summation term on the left hand side and solve for the summation term, then substitute back into (\ref{EqQuantSum}) to get
\begin{align}
S_{q}(k)=S_{v}^{\frac{2}{3}}(k)\left[\frac{\frac{f_{B}-f_{A}}{K}\sum_{k=1}^{K}S_{v}^{-\frac{1}{3}}(k)}{\sqrt{12}P}\right]^{2}.
\label{EqOptimalSqDiscrete}
\end{align}
Letting $K\rightarrow\infty$ in (\ref{EqOptimalSqDiscrete}) yields
\begin{align}
S_{q}(f)=S_{v}^{\frac{2}{3}}(f)\left[\frac{\int_{f_{A}}^{f_{B}}S_{v}^{-\frac{1}{3}}(f)df}{\sqrt{12}P}\right]^{2}
\label{EqOptimalSq}
\end{align}
which explicitly relates the optimal quantization noise shape to the signal noise shape.

Considering (\ref{EqOptimalSq}) in more detail, the squared term in brackets on the right hand side is a constant which is made smaller by increasing the power of the ADC.  Thus, the optimal quantization noise shape is proportional to $S_{v}^{\frac{2}{3}}(f)$.  Without the $\frac{2}{3}$ power, the optimal quantization noise PSD would be a fixed offset from the noise PSD regardless of the level of the noise PSD.  The $\frac{2}{3}$ power effectively shrinks the gap between the optimal quantization noise PSD and the noise PSD in low noise regions.  As such, while additional power in the ADC is allocated to low noise frequencies relative to high noise frequencies, the amount of additional power is constrained.

\begin{figure}[htb]
\centering
\includegraphics[width=8.6cm]{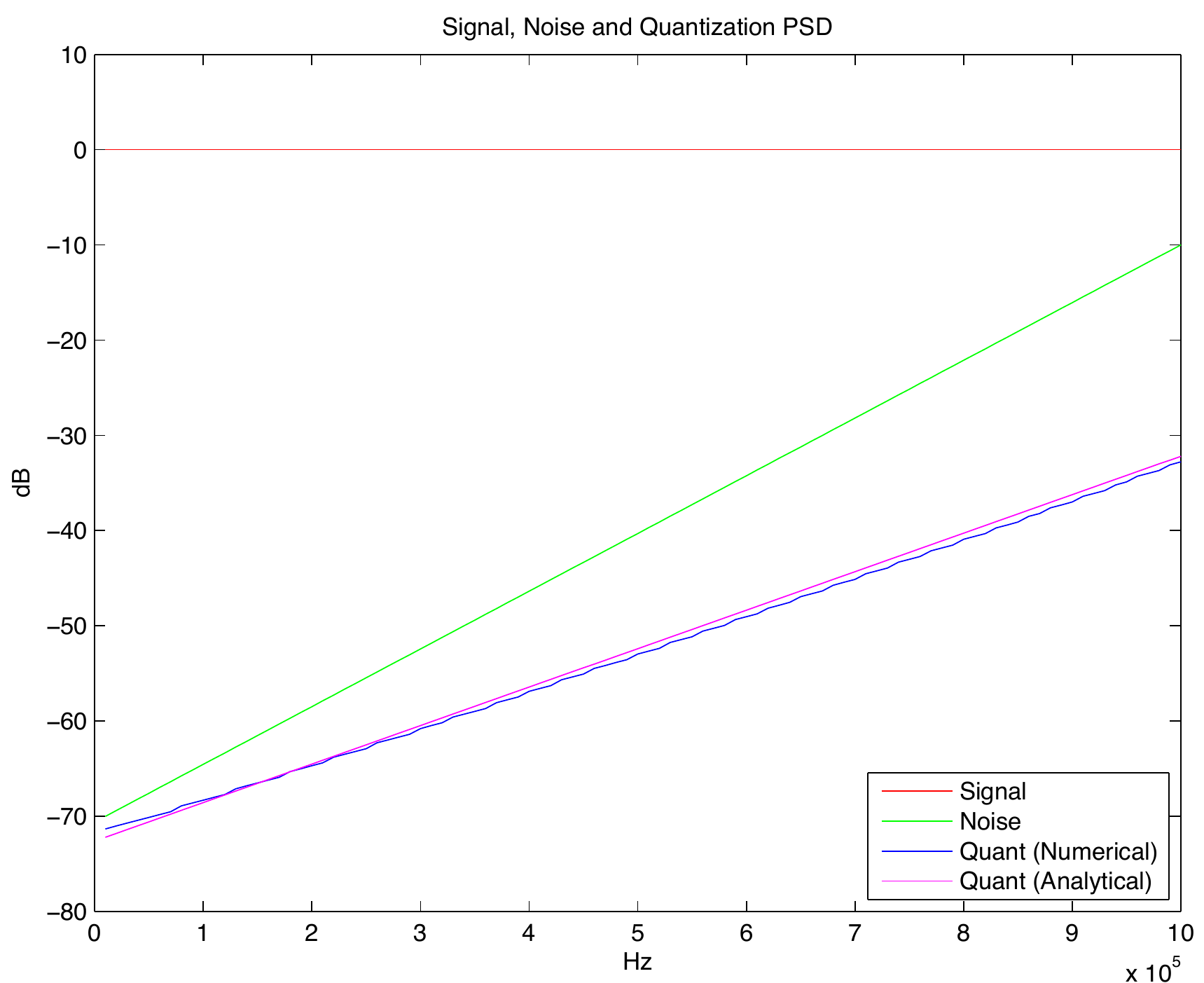}
\caption{A wireline style channel example optimal quantization noise PSD computed numerically (blue) and analytically (magenta) from (\ref{EqOptimalSq}) for maximizing information after the ADC with signal PSD (red) and noise PSD (green).}
\label{FigureQuantizationNoise1}
\end{figure}

\begin{figure}[htb]
\centering
\includegraphics[width=8.6cm]{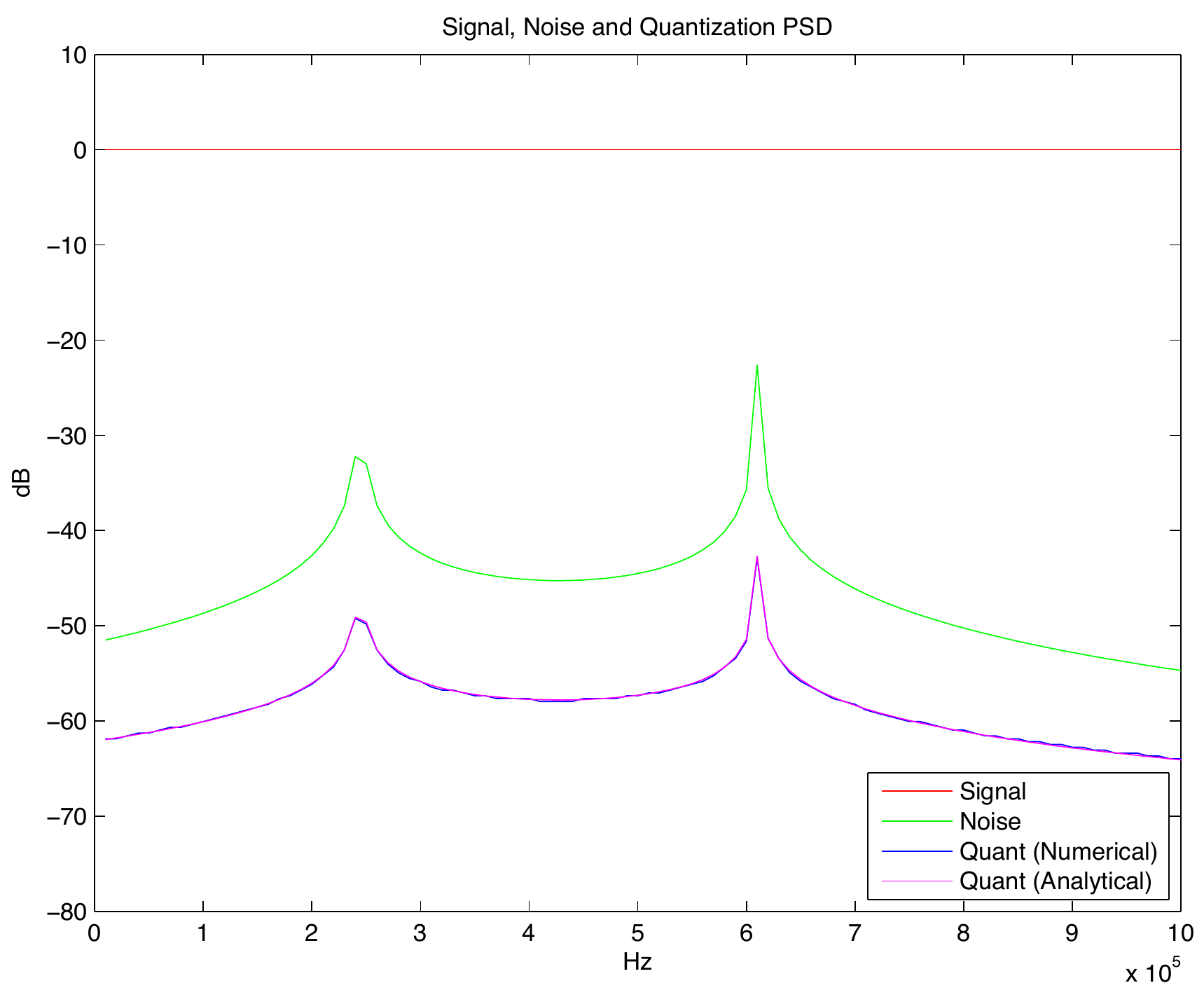}
\caption{A wireless style channel example optimal quantization noise PSD computed numerically (blue) and analytically (magenta) from (\ref{EqOptimalSq}) for maximizing information after the ADC with signal PSD (red) and noise PSD (green).}
\label{FigureQuantizationNoise2}
\end{figure}

Figs.\ \ref{FigureQuantizationNoise1} and \ref{FigureQuantizationNoise2} show examples of the optimal quantization noise PSD for maximizing information after the ADC.  Equation (\ref{EqOptimalSq}) was used to generate the analytical quantization noise curves.  The numerical quantization noise curves were generated by a stochastic search algorithm designed to minimize (\ref{EqInfoLoss}) given the power constraint (\ref{EqPowerConstraint}) and serve as a check on the theoretical result.

%
%

\section{Single channel ADC quantization noise shaping}
\label{SectionSingle}

\begin{figure}[htb]
\centering
\includegraphics[width=8.6cm]{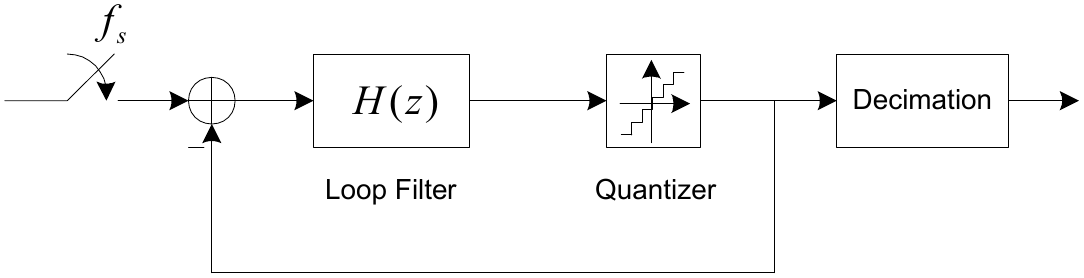}
\caption{A delta sigma ADC.}
\label{FigureDeltaSigmaAdc}
\end{figure}

Delta sigma ADCs achieve noise shaping through oversampling and a feedback loop with an embedded quantizer (see Fig.\ \ref{FigureDeltaSigmaAdc}).  Using a low pass delta sigma ADC as an example, the loop filter is designed such that the gain is large inside the signal band and small outside the signal band to allow the input signal and the analog feedback of the modulator output to match closely within the signal band.  Consequently, most of the signal difference at the summation node will be at higher frequencies and generate a shaped quantization error with it's power pushed outside the signal band.

A delta sigma ADC can be represented in the z domain by \cite{REF7}
\begin{align}
Y(z)=\text{STF}(z)X(z)+\text{NTF}(z)Q(z),
\end{align}
where $X(z)$, $Y(z)$ and $Q(z)$ are the z transforms of the ADC input, output and quantization error, respectively, and $\text{STF}(z)$ and $\text{NTF}(z)$ are the signal and noise transfer functions given by
\begin{align}
\text{STF}(z)=\frac{H(z)}{1+H(z)}\text{ and }\text{NTF}(z)=\frac{1}{1+H(z)}.
\end{align}

Note that using the structure in Fig.\ \ref{FigureDeltaSigmaAdc} shapes the quantization noise PSD as
\begin{align}
S_{q}(f)=\frac{\Delta^{2}}{12f_{s}}\left|\text{NTF}\left(z=e^{j2\pi
f/f_{s}}\right)\right|^{2},
\end{align}
where $\Delta$ is the quantization step size and $f_{s}$ is the sampling frequency.  Noise shaping can thus be achieved through the design of the loop filter $H(z)$.  Since it is possible to control the filter coefficients through adjusting feedback currents in the analog IC, noise shaping can be decided in the digital domain according to (\ref{EqOptimalSq}) and then realized in the ADC though controlling feedback currents.

\begin{figure}[htb]
\centering
\includegraphics[width=8.6cm]{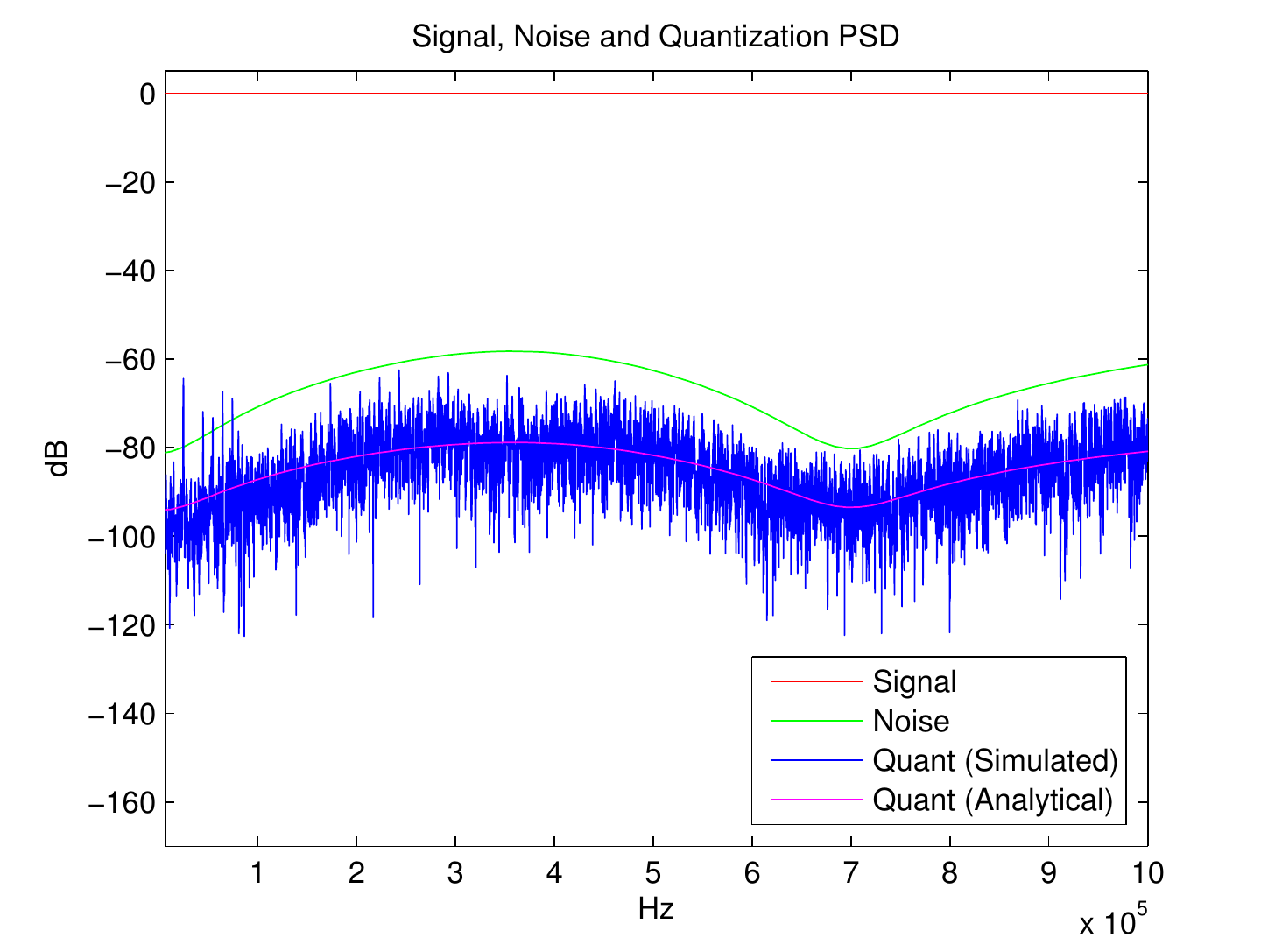}
\caption{A wireless style channel example optimal quantization noise PSD from the simulated delta sigma ADC (blue) and analytically (magenta) from (\ref{EqOptimalSq}) for maximizing information after the ADC with signal PSD (red) and noise PSD (green).}
\label{FigureExamples}
\end{figure}

As an example, a 4th order delta sigma ADC with an oversampling ratio of 12 was simulated with a loop filter $H(z)$ optimized to achieve quantization noise shaping as in (\ref{EqOptimalSq}) for the example where the channel has a shaped noise spectrum.  The resulting signal, noise, ideal quantization and actual quantization PSDs are shown in Fig.\ \ref{FigureExamples}.

%
%

\section{Multi channel ADC quantization noise shaping}
\label{SectionMulti}
Single channel ADCs, each able to optimally shape their quantization noise, can be combined to create a multi channel ADC using any of the traditional multi channel structures.

For the case of time interleaving, a set of $N$ individual ADCs with appropriate time offsets and matching can be combined to form an ADC with an overall quantization noise PSD shape that resembles a $N$x bandwidth expanded version of the quantization noise PSD of an individual ADC.

\begin{figure}[htb]
\centering
\includegraphics[width=8.6cm]{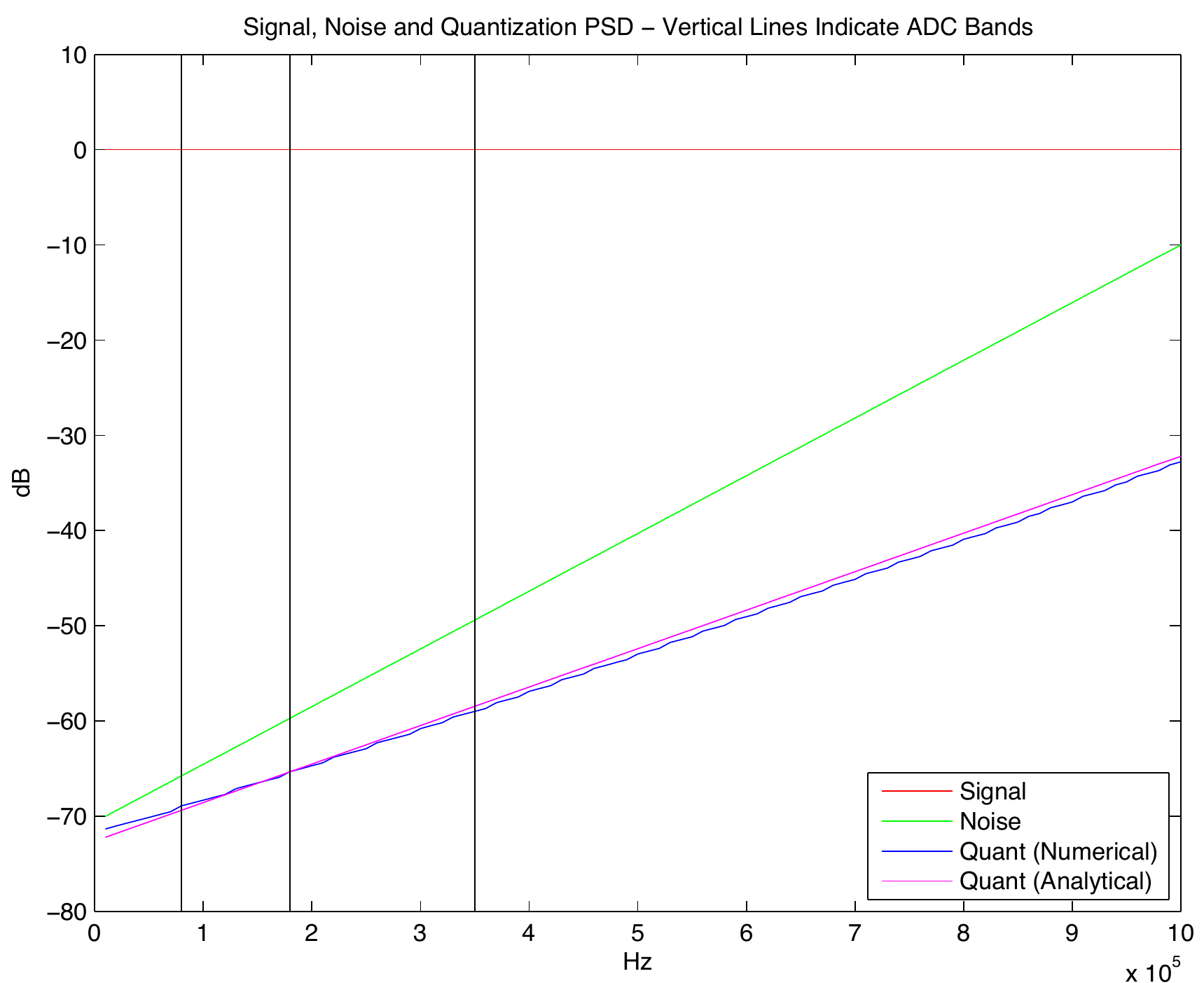}
\caption{A wireline style channel example.  Vertical black lines indicate the partitioning of the total bandwidth to the individual ADCs such that the power of each ADC is equal and the information after the multichannel ADC is maximized.}
\label{FigureMultiChannel}
\end{figure}

For the case of frequency interleaving, the total bandwidth can be divided into $N$ contiguous bands such that the power given in (\ref{EqPowerConstraint}) is equal for each ADC.  Note that this may result in an unequal distribution in frequency of the total bandwidth.  Using the wireline style system as an example, Fig.\ \ref{FigureMultiChannel} shows how the total bandwidth is split between $N = 4$ ADCs such that the power of each ADC is equal and the information after the multi channel ADC structure is maximized.

To simplify ADC design and combining, an additional constraint of equal or integer scale factors of bandwidth could included.

%
%

\section{Conclusions}
\label{SectionConclusions}

This paper derived the optimal quantization noise PSD shape to maximize the information content in a signal after an ADC with a power constraint.  It was shown that quantization noise shaping can be realized via loop filter design for a single channel delta sigma ADC and extended to common time and frequency interleaved multi channel structures.

%
%


%
%

\vspace{-1.9 in}

\begin{IEEEbiography}[{\includegraphics[width=1in,height=1.25in,clip,keepaspectratio]{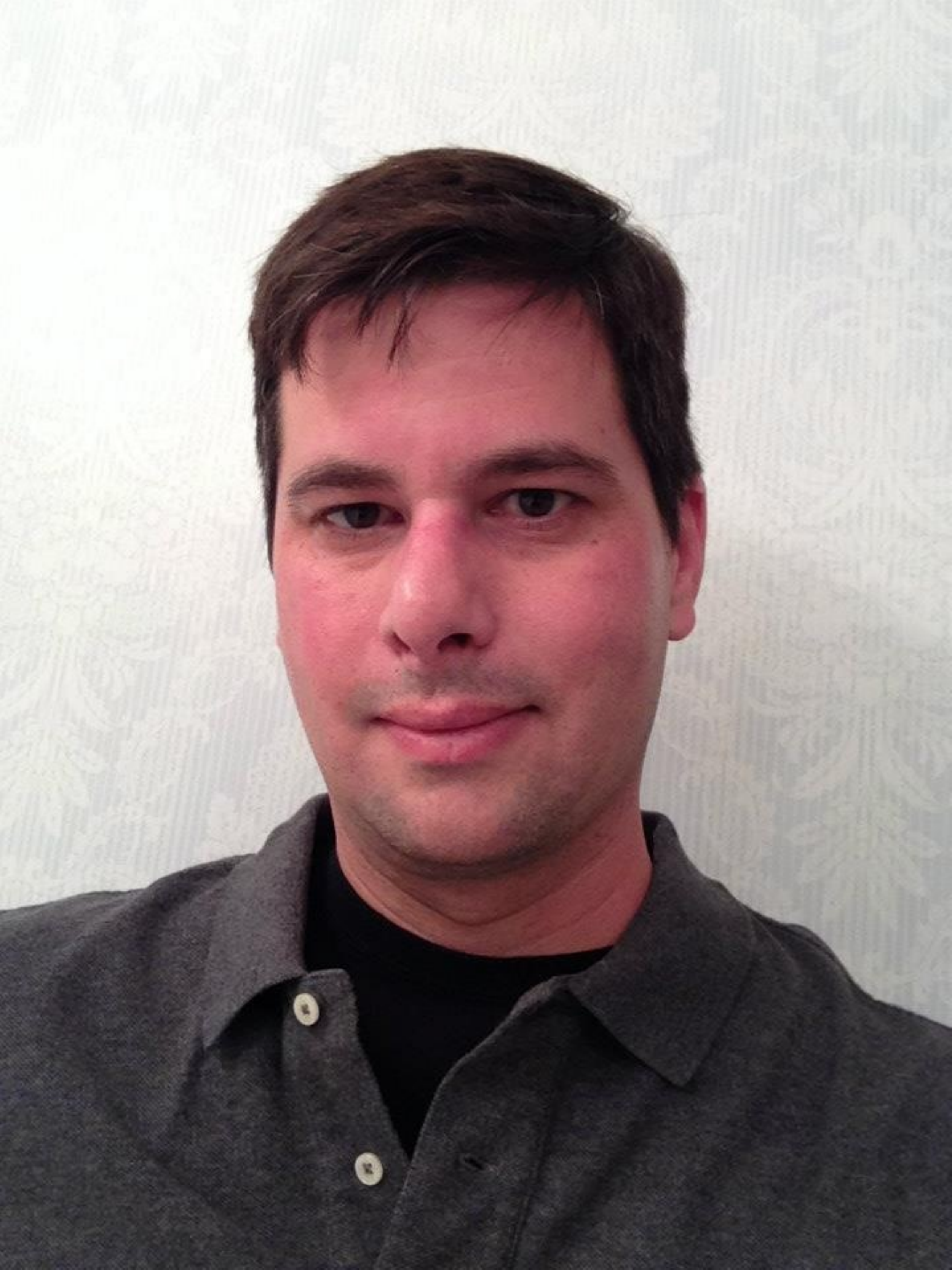}}]{Arthur Redfern}
Arthur J. Redfern received a B.S. in 1995 from the University of Virginia and M.S. and Ph.D. in 1996 and 1999, respectively, from the Georgia Institute of Technology, all in electrical engineering.  While at Georgia Tech he was supported by the Robert G. Shackelford fellowship from the Georgia Tech Research Institute and a Graduate Student Research Program fellowship from the National Aeronautics and Space Administration.  

Following his thesis work on data aided and blind equalization of nonlinear communication channels modeled by the Volterra series, Arthur joined the Systems and Applications R\&D Center at Texas Instruments where he currently manages the Signal Processing for Analog Systems branch.  His activities at TI have spanned the areas of ADC architectures and compensation, antenna tuning, PA compensation, speaker protection, touch screen controllers, wireless systems (DTV and BAN) and wireline systems (DSL and SerDes).  He has been granted over 20 US patents.

Arthur's hobbies include cars, poker and quantitative trading.
\end{IEEEbiography}

\vspace{-1.9 in}

\begin{IEEEbiography}[{\includegraphics[width=1in,height=1.25in,clip,keepaspectratio]{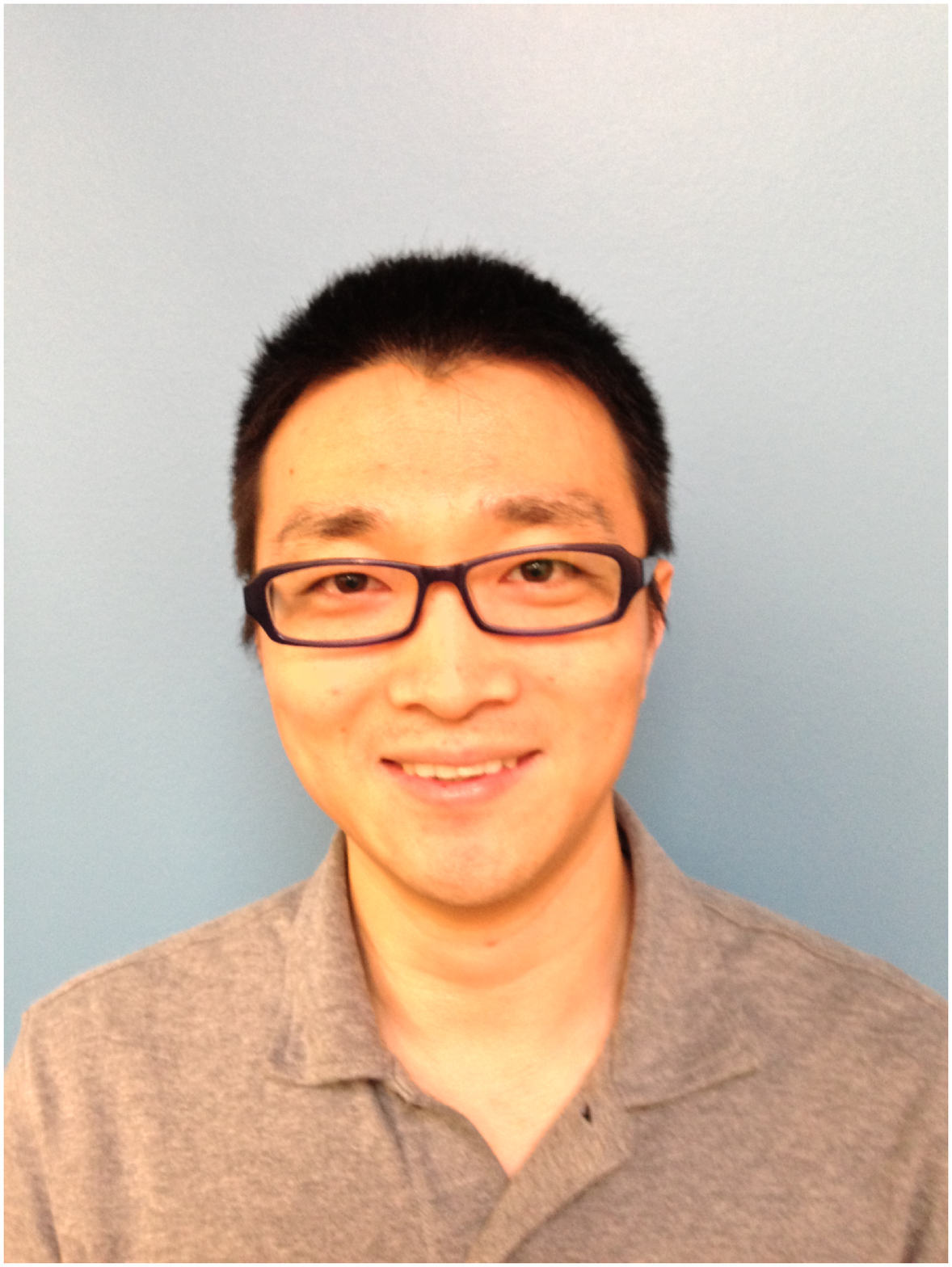}}]{Kun Shi}
Kun Shi received a B.S. in 2002 from the Beijing University of Posts and Telecommunications, M.S. in 2005 from Tsinghua University and Ph.D. in 2008 from Georgia Institute of Technology, all in electrical engineering.
 
Since January 2009, Kun has been with the Systems and Applications R\&D Center at Texas Instruments. His research interests are in the general areas of statistical signal processing, nonlinear systems and adaptive methods.
\end{IEEEbiography}


%
%


\end{document}